\documentstyle[psfig]{mn}

\newcommand{\alt}{\ \raisebox{-.3ex}{$\stackrel{<}{\scriptstyle \sim}$}\ }
\newcommand{\agt}{\ \raisebox{-.3ex}{$\stackrel{>}{\scriptstyle \sim}$}\ }

\newcommand{\degree}{$^{\circ}$}

\newcommand{\mum}{$\mu$m}
\newcommand{\by}{$\times$}

\title[Near IR star counts]{Near infrared star counts as a test of Galactic bar
structure}

\author[M.Unavane et al.]{
M.~Unavane,$^1$ Gerard~Gilmore$^{1,2}$\\
$^1$ Institute of Astronomy, University of Cambridge,
Madingley Road, Cambridge CB3 0HA, UK
\\
$^2$ Institut d'Astrophysique de Paris, 98bis Boulevard Arago,
F-75014 Paris, France
}

\date{\today}

\begin{document}
\maketitle
\begin{abstract}

We present survey data in the narrow-band L filter (nbL), taken at UKIRT, for
a total area of 277 arcmin$^2$, roughly equally divided between four regions
at zero Galactic latitude and longitudes $\pm$4.3\degree\ and $\pm$2.3\degree.
The 80 per cent completeness level for these observations is at roughly magnitude
11.0. This magnitude limit, owing to the low coefficient for interstellar
extinction at this wavelength ($A_{nbL}$=0.047$A_V$), allows us to observe
bulge giants. We match the nbL-magnitudes with DENIS survey K magnitudes, and
find 95 per cent of nbL sources are matched to K sources. Constructing
colour-magnitude diagrams, we deredden the magnitudes and find evidence for a
longitude dependent asymmetry in the source counts. We find that there are
$\sim$15 per cent and $\sim$5 per cent more sources at the negative longitude than at the
corresponding positive longitude, for the fields at $\pm$4.3\degree and
$\pm$2.3\degree respectively. This is compared with the predictions of some
Galactic bar models. We find an asymmetry in the expected sense, which
favours gas dynamical
models and  the recent deconvolution of surface photometry data (Binney et al.
1991; Binney, Gerhard \& Spergel 1997), over earlier
treatments of photometric data
(e.g. Dwek et al. 1995).
\end{abstract}

\section{Introduction}

An extremely dense stellar cluster and probable bar dominate the central
kiloparsec of the Galaxy. This cluster may be the remnant of a core about
which the galaxy grew or may be the product of a long-lived bar in the disk
feeding gas into continuing central  star forming regions. Neither its
history and nature, nor is its relationship to the bulge, halo and disk are
well known.

As an example, the cluster changes its luminosity density profile
by 2 in the power law index in some unobserved region between the
central few arcsec and the optically observable region some degrees
away.

How and where? And are more complex spatial distributions possible?
For example, in M31, the nearest similar spiral, the
central region shows two luminosity maxima, neither of which
corresponds to the centre of the larger scale gravitational potential,
or is understood. (van der Marel et al. 1997)

In practice, to observe the central regions of our own Galaxy, because of the
high extinction in the disk in the line of sight, it is necessary to work at
infrared wavelengths, at which the dust is more transparent, or to use
wavelengths very different from optical wavelengths where dust does not
impede out view (e.g. radio frequencies). In the former case, we use stars as
tracers of structure, and in the latter case, we trace gas, or young stars
with associated HII regions.

In doing this, many studies have suggested the existence of a central
kiloparsec-scale bar. Gas dynamical data has been used by Binney et al.
(1991) and Blitz \& Spergel (1991) to derive the existence of a bar with its long
axis pointing towards the first quadrant ($\ell>0$). In terms of near
infrared surface photometry, the measurements made by the COBE satellite have
been the most extensive. The DIRBE instrument on board mapped the whole sky
at near to far infrared wavelengths from 1.25\mum\ to 240\mum\ using the
DIRBE with a beam of 0.7\degree\ \by 0.7\degree.

This data has been used, among others, by Dwek et al. (1995) whose best fit
model suggests a boxy-bar in the centre of our Galaxy with an inclination of
about 19\degree\ to our line of sight, the long axis again being in the first
quadrant. The latest and most sophisticated use of the DIRBE data has been by
Binney, Gerhard \& Spergel (1997), who do a full three-dimensional deconvolution of the
apparent light intensity and dust distribution. Binney, Gerhard \& Spergel (1997) also
find a bar with its long axis in the first quadrant, inclined at about
20\degree. This is near the limit of what is possible with surface photometry
data of low spatial resolution, as demonstrated recently by Zhao (1997).

Other investigations have used tracer populations to constrain the nature of
the bar - for example, Weinberg (1992) and more recently Nikolaev \& Weinberg
(1997) use IRAS source to confirm that there is a large scale asymmetry in
their distribution, consistent in geometry with a bar-like structure.

Any bar-like structure will be centrally concentrated, and a relatively small
survey concentrating on sources near the centre will have enough statistical
weight to show up longitude dependent asymmetries, if they exist. The only
moderately sized, multicolour, higher resolution survey, covering about
$2^\circ \times 0.5^\circ$ around the Galactic centre, is by  Catchpole,
Whitelock and Glass (1990).  It was performed in the J, H and K bands up to a
limiting magnitude of K=12.  While almost the entire J map was dominated by
heavy interstellar extinction, those at H and K show progressively more detail
of the inner region. They show clear changes in spatial structure for different
populations, suggesting that analysis of low resolution data will necessarily
be problematic.

For the galactic centre this means that M and late K giants can be
reached in K but not at the shorter wavelengths as the extinction will
be too strong (up to 5 magnitudes in J [Catchpole, Whitelock \& Glass 1990]
leading to an expected apparent J magnitude of $\approx
16.5$~mag; [Wainscoat et al. 1992]). We also expect that
essentially all I and most J objects seen in the plane will be disk objects.

Recently, a high resolution survey in near infra-red bands called DENIS (Deep
Near Infra Red Southern Sky Survey) has been initiated. It aims to map the
whole of the southern sky in I,J and K with 3 arcsecond pixels. (Epchtein
1997; Fouqu\'e et al. 1997). The calibration and flat-fielding of images is
reported in Borsenberger (1997), and the source extraction and
astrometry that we use come from the accompanying paper 1 (Unavane et al.
1997). In that paper we reported techniques for removing the effects of
foreground disk asymmetries in the colour-magnitude diagrams, and used a
statistical approach to understanding asymmetries. The usefulness of the three
wavebands was somewhat limited since only K reliably penetrates the densest
dust regions in the plane towards the centre of the galaxy. A source by
source dereddening - necessary because of the high spatial frequency  structure
in extinctive clouds - was not possible.

Here we present complementary nbL data taken at UKIRT, which we match to DENIS K
data. A source by source dereddening is carried out, from which we deduce the
existence of central longitude asymmetries consistent in direction with
some Galactic bar models.

We present first the high-quality 1996 data, and subsequently a brief
description is given of the problematic 1995 data.

\section{The observations}

\begin{figure}
\hspace{7cm}
\label{showfields1}
\end{figure}

The locations of the regions observed in nbL are indicated in figure
\ref{showfields1}. Also indicated on this figure are the DENIS rasters for the
region. The region enclosed by the dotted box shows the set of UKIRT
observations made in July 1995, which were severely affected by a fault in the
data acquisition software. The regions observed in July 1996 achieved the
expected magnitude limits and were reduced using standard techniques. For
comparison, the beamsize of the COBE/DIRBE survey (0.7\degree) is shown in the
same figure.

\subsection{Why nbL (3.6\mum) ?}

\begin{figure}
\hspace{2.5cm}
\label{kldiag}
\end{figure}

\begin{figure}
\hspace{7cm}
\label{Lfilter}
\end{figure}

From the colour magnitude diagram in figure \ref{kldiag}, it can be seen that
by far the brightest sources are the late type III sources (i.e. M giants),
which reach absolute L magnitudes of $\sim$ -7. If we assume the distance to
the centre of the galaxy to be 8kpc, this corresponds to an apparent
magnitude at that distance of 7.5.  The extinction coefficient in nbL
(3.6\mum) is only $\sim$40 per cent of the value in magnitudes present at K, and
$<$5 per cent that at V. Thus, using an estimate of  1.9 magnitudes of extinction
per kpc in the disk at visual magnitudes (Schmidt-Kaler, 1976), the above
magnitude is increased to about 8.4. Assuming a limiting magnitude of 12 in
the nbL observations, this suggests that source as faint as $m_{L}$ = -3.4
should be detected (i.e. all bright K giants and all M giants should be
detected).

In addition, as explained in the next paragraph, this wavelength regime is a
bridge between shorter wavelength surveys (I,J,K) and longer wavelength
surveys (7\mum, 15\mum) which are being carried out near-simultaneously.

Furthermore, and a not unimportant factor when observing from the surface of
the Earth, the atmosphere has a transparent window at this wavelength.

\subsection{Choice of fields for observation}

Part of the mission of the ISO spacecraft, launched in November 1995, was to
map selected regions of the Galactic Plane at 7\mum\ and 15\mum. The
project, called ISOGAL, has provided and will continue to provide, high
resolution (6 arcsecond pixels) images which penetrate deep into the Galactic
Plane, by observing at these wavelengths where extinction by dust is very small.
The choice of nbL fields is made so as to coincide with some ISO fields, since
the amount of information about an astronomical source can be increased
manyfold with colour information.

Furthermore, the DENIS survey is producing images in I,J and K, for the whole
of the southern sky, which includes all of the Galactic Plane within several
tens of degrees of the centre. (see Epchtein 1997; paper 1). The nbL
observations also have full coverage at these DENIS wavelengths, again
allowing invaluable colour information to be deduced.

Also, following on from paper 1, we choose  fields at equal and opposite
longitudes at a given latitude to test for longitude asymmetries in
the inner disk/bulge. The expected maximum contrast for most Galactic bar
models falls between 4\degree\ and 6\degree\ in longitude - fields for
observations in this
region would be best at discriminating models.

Furthermore, to avoid corresponding positive and negative longitude pairs with
very different extinctions in the line of sight, we choose fields with similar
intergrated luminosities in the J and K bands, and hence in (J-K) colour also.
These wavelengths sample dust extinction
in the line of sight in the disk, and it is to
minimize these asymmetries that we make our choice. We base this on the
COBE/DIRBE maps by choosing equal and opposite longitude pairs which most
nearly show the same fluxes and colour. Figure \ref{dirbecont} shows the J,K
and (J-K) contours for the central few degrees of the Galactic bulge. The
chosen fields for observation are marked by asterisks, and some
contours are emphasized to show that they pass as nearly as possible, given
the various other constraints, through these fields.

\begin{figure}
\hspace{7cm}
\label{dirbecont}
\end{figure}

\section{The 1996 observations}

An enlargement showing the pattern of the 1996 observations is indicated in
figure \ref{showfields2}. The area of each field was calculated by noting the
perimeter points and applying the formula for the area of a region bounded by
$n$ given, ordered perimeter points $(x_i,y_i)$ (where $(x_{n+1},y_{n+1})
\equiv (x_1,y_1)$):

$$ A = \frac{1}{2} \sum_{i=1}^{n} \left| \begin{array}{cc}
                                            x_i   &   y_i   \\
                                          x_{i+1} & y_{i+1} \\
                                         \end{array} \right| $$

The correction for spherical geometry is negligible for fields of this size.

\begin{figure}
\hspace{2.5cm}
\label{showfields2}
\end{figure}

\subsection{At the telescope - data acquisition}

The observations were carried out at UKIRT during the
first half of each night on 3rd and 4th July 1996, during photometric
conditions. Cloud coverage prevented further observations.

 The instrument used was IRCAM3 (Infra Red
CAMera). IRCAM3 is a cooled 1-5\mum\ camera with a 256 $\times$ 256 InSb array.
The basic plate scale is $\sim$0.286 arcsec per pixel, giving a maximum field of
73 arcsec. Standard J,H,K, nbL, L' and narrow band M photometric filters are
available for use. In these observations, the nbL filter was chosen (narrow band
L). The central wavelength of this filter is 3.60\mum, with a FWHM of 0.06\mum.
The profile is shown in figure \ref{Lfilter}.

The mode of observation, which had been tested in the previous run in July 1995
(see below), was adopted. The mode allowed us to achieve a limiting magnitude of
nbL $\sim$ 12-13. A major difficulty, and one which becomes more difficult as
the wavelength is further increased above 3.6\mum, is the very high sky
background level, caused primarily by a blackbody-type emission from the sky,
upon which are superimposed sky emission lines.

For example, the peak pixel value of an 8th magnitude source is no more
than 140 counts in the typical exposures that we use, over a background count of
about 40000. This is a signal of only a few parts in a thousand, thus requiring
a very precise sky subtraction. As can be seen,
a long stare mode observational technique would immediately saturate the
detectors, and no signal from the sources would be seen.

The adopted method of observation was as follows:

$\bullet$ Each {\em raster} of observation consisted of a slightly-overlapping
grid of 3 $\times$ 3 {\em images}, and was preceded by one dark frame.

$\bullet$ Each of these {\em images} is the sum of 167 exposures, each of 0.12
second duration. The short duration of each individual exposure is essential to
avoid  saturation of the detector. Hence, the total exposure time per image was
20.04 seconds.

The individual exposures are not delivered, but are coadded by an on-line
 transputer array
(ALICE) and
the final image is delivered.

The time taken for the observations associated with each raster was about 10
minutes, after taking account of the overhead times in multiple detector
readouts, and telescope slewing.

Furthermore, at regular times during the observing period (in our case, after
every 3 rasters) a standard star raster was taken, which consisted of 5 images,
each image taken in just the same way as above, and an associated dark frame.
The chosen standard star was HD161903, which has an nbL magnitude of 7.00,
at $(\alpha_{2000},\delta_{2000}) = $ (17$^h$45$^m$43.3$^s$, -1$^{\circ}$47'34")
or $(l,b)$=(23.64, 13.73).

The observing pattern is indicated in the sketch in figure
\ref{ukirtraster}. The resulting data taken away from the telescope thus
consisted of several rasters, each containing 9 images, and a dark frame
associated with each one. In addition, several standard star rasters,
consisting of 5 dithered images and associated dark frame, are also taken.

\begin{figure}
\hspace{2.5cm}
\label{ukirtraster}
\end{figure}

\subsection{At the computer - data reduction}

A standard technique for reducing this data was employed. Below is the
procedure adopted for each raster:

\begin{enumerate}

\item  Dark subtraction - from each image in a raster, the dark frame was
       subtracted.

\item  A flat field image was created by median-filtering the set of 9 object images. (5 images in
the case of the standard star) The technique of using a "sky" frame devoid of sources fails in these
crowded Galactic centre regions. The flat field image is divided by the median of all the pixel
values in it, to normalize it.

\item Flat fielding - the dark subtracted images are divided by the normalised
      flat field image.

\item Pretty picture production - the 9 images can be mosaiced to form a pretty
      picture for display purposes.

\item Source extraction - all images were filtered using a moving block median,
and sources more than 2.3$\sigma$ above the background noise were flagged, by use
of the SExtractor program (Bertin \& Arnouts 1996). The photometry was then carried out on
the unfiltered images by using a 7 pixel radius aperture for the source, and an
annulus between 7 and 11 pixels from the source centre for the sky background to
be subtracted from it.

\item Airmass correction - using the airmass information in the header, and the standard value for
the airmass correction at the UKIRT site (0.09 magnitudes per airmass), all the raw magnitudes were
corrected.

\item Absolute photometry - the corrected standard star photometry was used to determine the offset
between instrumental and true magnitude. The correction was applied to all extracted sources.
A very stable mean zero point of 18.78$\pm$0.03 magnitudes was derived.

\item Approximate astrometry - The positions of sources were calculated assuming that the pixel scale
was 0.286 arcseconds per pixel and by using the header information about the
image centres.

\item Absolute astrometry - see below

\end{enumerate}

Table \ref{tab2} shows the numbers of sources found in each of the fields.

\subsection{Photometric error and completeness}

The random uncertainty in magnitude was assessed empirically by
comparing the magnitude of the same object when it appeared fully
twice (or more) in neighbouring images.
Figure \ref{Lbandmagscatter} shows, in the upper panel, the difference
in magnitude between measurements of the same object from overlapping images.
A clipped gaussian was fitted to the distribution
in several magnitude ranges for this diagram, to give values $\sigma_{m_1-m_2}$
representative of this random scatter in magnitude differences.
 This $\sigma$ represents the
standard deviation of the difference of two like distributions. Hence the
standard distribution for individual magnitude measurements is given by
$\sigma_m = \sigma_{m_1-m_2}/\surd2$. The result
is shown in the lower panel of figure \ref{Lbandscatter}.

Completeness was assessed by the addition of artificial stars to the images. Two
images were used - one from the first of the two nights of observation which was
judged by eye and by image background statistics to be particularly noisy after
reduction, and one from the second of the two nights which was judged to be
particularly clean. The ellipticity, orientation, and full-width at half-maximum
values for the stellar sources were averaged for each image, and used to create
artificial stars with these same parameters. In each case, magnitudes for
artificial stars ranging from 10.00 to 13.50 in 0.25 magnitude intervals were
used, with never more than 25 artificial stars being added to each image. Hence,
a total of 30 modified images were created (15 for each original image), and
were each reduced in the same way as described above for the untouched images.

The fraction of retrieved images for each magnitude range is shown
graphically  in figure \ref{Lcompleteness}. The fall below 90 per cent completeness
occurs for both images at about magnitude 11.0, with a difference of no more
than 0.25 magnitudes between the two images. Unlike the DENIS images (paper
1), there is essentially no problem with crowding in these fields, since the
pixel size used is small (0.286 arcsecs) and the sources are seldom within a
few arcseconds of one another. As a result, there is little difference in
completeness levels if the positional tolerance of matches of recovered
images is made tighter, or looser.

\begin{figure}
\hspace{2.5cm}
\label{Lbandmagscatter}
\end{figure}


\begin{figure}
\hspace{2.5cm}
\label{Lcompleteness}
\end{figure}

\subsection{Absolute positions}

An absolute reference frame was established by the use of a
source extraction of a plate scan matched to GSC stars. (See paper 1). This
optical reference frame cannot be applied directly to the L band images, because
in the directions of high extinction towards which the fields lie, it is
impossible to establish with any certainty which L sources in the images (which
usually number at most a dozen) correspond to optical sources. Moreover, despite
the generous overlaps between images, expected to enable a precise mosaicing to
be carried out, there remain problems in very many of the images due to random
telescope motions which cannot be corrected because of a lack of common sources
in the overlaps.

The method finally employed to fix the positions involves the use of the DENIS
results in the same regions. For each of the nbL fields, we have corresponding
DENIS I,J and K images (see figure \ref{showfields1}). A "step-up" method can be
used. The sufficient similarity in sources between 0.6\mum\ and 1.25\mum\ can be used
 to match
up optical and J images. This puts the J images on the same astrometric
reference frame as the optical images. From there, the K images can be mapped
onto the J images by source matching, and finally the nbL sources can be matched to
the K sources.

This final procedure was carried out by eye for each of the 288 images in the
1996 observations.

\section{The 1995 Observations}

The observations were obtained during the first1996 half of the nights of
7,8 and 9 July 1995, at UKIRT using the narrow-band L filter. Reference is made
to these observations in an earlier article (Unavane \& Gilmore 1997).

Five regions were imaged in the same manner as for the subsequent 1996
observations. The pixel size was 0.286 arcseconds, and the exposure time for
each image was a total of 20 seconds.  The flat fielding was not
carried out in the standard way, due to technical difficulties which plagued
the acquisition transputer array. Furthermore, two out of the three
half-nights allocated were marred by thin cloud. The electronic noise in the
image consisted of both systematic pixel value shifts and random noise. Each
image was inspected for the nature of the systematic shifts, and an attempt
was made to interpolate over such rows and columns using "good" neighbouring
data. The small pixel size helped to retain adequate resolution nevertheless.
The density of cloud was empirically corrected for by finding a correlation
between the sky background and the varying apparent magnitude of the standard
star.

The magnitude zero points were taken from observing the standard star HD161903
as in the 1996 observations.

Perhaps unsurprisingly, the completeness and photometric precision were below
expectation. A limiting magnitude of about 9.5 was achieved for the nights
when electronic faults dominated uncertainties, and of about 11.0 when only
cloud dominated the uncertainties. Random
photometric scatter was estimated by comparing the magnitudes obtained from
neighbouring images. Scatter of up to 0.5 magnitudes is present in most of the
observations. It is interesting to note that the empirical method of
correcting for cloud, though it may have severe systematic uncertainties
associated with it, seems to work better than the method for correcting
electronic noise.

\begin{figure}
\label{scatter1995}
\end{figure}

The locations, and labels for the regions are given in table 1. Also included
are the area of the fields, and an indication of the relative numbers of
sources per unit area in the fields. The number counts are shown in figure
\ref{numbercounts}.

Apart from noise at bright magnitudes due to low number statistics, the
populations appear to follow similar trends, until nbL$>\sim$8.5 where the
fields A, C and E, show a significant excess. This may plausibly be due to
regions of lesser extinction in these fields. The final column in Table 1 gives
the number count as log$_{10}$(number of sources/magnitude/deg$^2$)
interpolated to L=7.5  after a linear fit between 6.0 and 9.0. The central
field, A, has a significantly higher count, by about a factor of two, than the
other fields.

\section{Analysis}

\subsection{Extinction coefficient}

The extinction coefficient associated with this filter may be calculated by
convolving its profile with the reddening curve. Using the parametrization of
the reddening curve of Mathis (1990) between 1.0\mum\ and 3.8\mum\ in paper 1
( $ \frac{A(\lambda)}{A(J)} = 1.484 - 5.60109 x +8.395624 x^2 -4.5947083 x^3 $
where $x=log_{10}(\lambda/\mu m)$ ), we find that the coefficient is given by
$A_{nbL}$=0.047$A_V$. Convolving in the UKIRT sky makes no appreciable
difference, since this part of this region of the infrared spectrum is
relatively free of strong lines. Compared with DENIS-K, for which the value is
$A_K$=0.112$A_V$, we see that the extinction is only $\sim$ 40 per cent of the
K-value in the band nbL, and $<$5 per cent of the value in the visible.

\subsection{The 1995 observations}

We note that the 1995 fields B and C, which are at longitudes of $1^{\circ}$
and $-1^{\circ}$ respectively, show a significant difference in counts, with
the field at $\ell=-1^{\circ}$ being more populous by a factor of $\sim$1.6,
and that the fields D and E, which we would expect to show number counts
falling off steeply, as $R^{-1.8}$ (King, 1989), have a very similar number
count.

In both cases, {\it extinction} along the line of sight may play a crucial
role. Calculation shows that in the former case, in order to make the regions B
and C display comparable counts would require a relative shift in L of
$\sim$0.7 magnitudes, which corresponds to a difference in extinction, in the
visual band, between the two lines of sight of $\Delta A_V\sim12$. For the
latter case, on integrating along the line of sight the expected $R^{-1.8}$
bulge distribution, we would predict number counts different by a factor of
$\sim$2.3, corresponding to $\Delta A_V\sim11$.

These values of $\Delta A_V$ of $\sim$12 correspond to differential
reddening $\Delta (J-K)$=2.0 and $\Delta (K-L)$ =0.6, using the values of
$A_X$
derived above. These values are large enough to be tested by multicolour
observations in the various fields, in order to disentangle the effects of
reddening from those of stellar distribution.

\subsubsection{Combining with DENIS-K}

We use the photometric and astrometric reductions of DENIS data described in
paper 1. Only data in regions B and E contained sufficiently reliable images
suitable for matching to DENIS data. The matching between sources led to only
26 per cent and 18 per cent matches in regions B and E respectively, reflecting the poor
quality of data acquired in this run. Compare this to the 1996 data where
\agt 95 per cent of L sources are matched with K sources (see later). No attempt was
made to understand the selection effects in this matching.

The resulting colour magnitude diagrams are shown in figure \ref{kl1995}.

\begin{figure}
\label{kl1995}
\end{figure}

By tracing a fiducial colour-magnitude (K-nbL)-(K) locus for K and M giants
placed at a distance of 8 kpc (Garwood \& Jones 1987), we can estimate the
extinction to the sources. This method assumes that the objects are all bulge
objects, and that they are all giant stars. A rough calculation based on a
recently derived photometric model of the galaxy by Binney, Gerhard \& Spergel
(1997)
suggests that inward of 3kpc from the centre, no more than 12 per cent of sources
are disk sources (see paper 1).  Each source can be traced back along a
reddening vector to the fiducial locus.

Figure \ref{kl1995} suggests a near constant extinction for region E of
$A_V$ between about 10 and 15, bearing in mind the large scatter in the
L-band photometry for this field. In the case of region B, however, we
clearly see a larger spread of extinction values between $A_V=5$ and
$A_V=35$. The fieldsize for these K and L sources is only some tens of
arcmin$^2$, yet even on that scale, there is a wide range of extinction
present. Mapping this extinction suggests an essentially random
distribution in this region.

This is consistent with the known molecular cloud distribution near this
region.
(Liszt, 1988). Sgr D, centred near $(\ell,b)\sim(1.1^{\circ},-0.1^{\circ})$,
has a concentration of molecular gas in it vicinity, as does Sgr B (centred
near
$0.6^{\circ},-0.1^{\circ}$), and the overlap of these molecular cloud regions
undoubtedly leads to the surprisingly patchy extinction pattern.

Nevertheless, these admittedly poor quality observations serve to indicate
that although the technique of dereddening sources by using multi-band
observations is conceptually simple, the direction of the fields must be
carefully chosen to avoid regions such as region B which have very different
reddenings on scales of a few tens of arcminutes.

For the 1996 observations, fields were chosen which were further out, to avoid
the worst Galactic centre molecular cloud complexes, and allow a better
characterisation of the stellar distributions.

\begin{figure}
\label{numbercounts}
\end{figure}

\begin{table}
\begin{center}
\small
\end{center}
\label{tab1}
\normalsize
\end{table}

\subsection{The 1996 observations}

Figure \ref{Lnumcnts} shows the resulting number counts for the four regions
at $\ell=$+4\degree, $\ell=$+2\degree, $\ell=-$2\degree, and $\ell=-$4\degree
. The raw counts in each case are given in panel (a). It is clear that there
is no large difference between number counts at like positive and negative
longitudes. Panel (b) shows the same data corrected for incompleteness
according to the completeness levels derived in the previous section. The
completeness level used was equal to the mean of the completeness levels
derived in the two separate experiments indicated in figure
\ref{Lcompleteness}.


\begin{figure}
\hspace{2.5cm}
\label{Lnumcnts}
\end{figure}

Absolute astrometry  and cross matching to
DENIS K data were carried out as described above. In establishing the
absolute positions of the sources, a cross-matched catalogue of the L and K
sources was made. More than 95 percent of sources detected in L were matched
by sources in K.

\begin{table}
\begin{center}
\small
\end{center}
\normalsize
\label{tab2}
\end{table}

\section{Reddening corrections}

The almost complete matching between K and L sources suggests that the K band
observations penetrate as much of the dust as do the L band observations. Since
the extinction coefficient in L is less than half that at K, we deduce that
the central bulge, which dominates the source counts, is reached in both
bands.

\begin{figure}
\hspace{2.5cm}
\label{drawcm}
\end{figure}

Plotted in figure \ref{drawcm} are the resulting colour-magnitude diagrams.
Shown in each case, as dotted lines, is an unreddened
giant branch displaced to a distance
of 8kpc. The differences in reddening along the line of sight to these fields,
and within the fields are clearly seen. The fields at $\ell=$4\degree\ and
$\ell=-$4\degree\
suggest that extinction is greater by some 0.6 in K-L (9 in $A_V$) at $\ell=$4 by
comparison of the location of the observed giant branches. The fields at
$\ell=$+2\degree\ and $\ell=$-2\degree, on the other hand, show more nearly similar observed
mean giant branch colours.

A simple approach to assessing number counts unaffected by reddening is to trace
back points in the (K)-(K-nbL) diagram so that they lie on the fiducial giant
branch. Thus, for each source, a value $A_{K-nbL}$ is established by tracing back
along the reddening line until the giant branch locus is reached. This value of
$A_{K-nbL}$ is converted to a value $A_{nbL}$ using $A_{nbL} = 0.723 A_{K-nbL}$.

These dereddened number counts are shown in figure \ref{modcnts}. Points at the
incomplete, fainter magnitudes are included weighted by the completeness function.
All points which would give a negative value for $A_V$ are rejected. These account for \alt
10 per cent of the sources, and can be attributed to photometric and/or matching errors.
The results for the mid-magnitude bins are shown in table \ref{tab3}.

\begin{figure}
\hspace{2.5cm}
\label{modcnts}
\end{figure}

\begin{table}
\label{tab3}
\end{table}

\section{Bar Structure}

A bar-like structure in the central regions of the Galaxy would show up as
a positive longitude /negative longitude asymmetry. Most bar models to date
agree roughly as to the orientation of the bar  in suggesting that the
nearer end is in the quadrant $\ell>$0\degree.  This means that the counts
are  higher at small {\em negative} longitudes ($|\ell |< \sim$6\degree)
than at the corresponding positive longitudes due to the geometry of the
viewpoint. For the Dwek et al. (1995) models the percentage difference in
surface luminosity from the bar, assuming no extinction, reaches a maximum
value in this region at $\sim$ 4\degree, when the flux from negative
longitudes is greater by $\sim$ 30 per cent than at the corresponding positive
longitude. The value is reduced to about 20 per cent greater at
$\ell=\pm$2\degree.

Figure \ref{baras2} shows the values of table \ref{tab3} value superposed on the
diagram of bar asymmetries in number counts expected  for the various bar
models. Note that the asymmetries which are shown in figure \ref{baras2} are not
direct photometric asymemtries, but rather the asymmetries resulting from a
numerical integration using the models derived from photometry. \footnote{For the
Dwek et al. (1995) models, we use their `best-fit' models E3 and G2, which come
from the fit to 2.2$\mu$m surface intensity, with an imposed radial
2.4\,kpc cut-off.} They are number
count contrasts expected for tracer populations. For our observations, which do
not extend beyond $\ell$=$\pm$4.3$^\circ$, the difference in distance modulus to
the two sides of the bar that we see is typically less than a few tenths of a
magnitude, and we use the nbL-band number counts as a tracer population.

Note also that the photometric models we use for comparison are derived from
observations at higher latitudes than the ones at which we are observing, where
the reddening is lower, which inevitably means that they are more reliable
there. Derivation of bulge structure from these data in the mid-plane, where
uncertainties in the removal of effects due to dust are greater, are less
reliable, which emphasises the value of resolved number count
studies such as this one.

In all of these directions, close as they are to the disk plane, there will be
contamination by disk sources. We make an estimate based on the Galactic model of Binney, Gerhard \& Spergel
(1997) for the contrast between disk and bulge in the inner disk. For sources
inward of 3\,kpc from the centre, observed from the sun, and for longitudes of
$<$5$^\circ$,  the number count contrasts vary strongly only with latitude
(because of the thinness of the disk), and are typically 10--12 per cent for
$|b|<0.1^\circ$. The actual central asymmetries are thus likely to be greater
than those seen here (by about 2 and 4 percentage points at $\ell$=$\pm4^\circ$
and $\ell$=$\pm2^\circ$ respectively) due to the diluting effect of the disk. We
thus find that we favour the photometric model of Binney, Gerhard \&
Spergel (1997) and the dynamical models of Binney et al. (1991) and Blitz
\& Spergel (1991) over the Dwek et al. (1995) photometry.

\begin{figure}
\label{baras2}
\end{figure}

L band imaging clearly has the potential to penetrate regions of severe
extinction - visual extinction of $A_V=30$ magnitudes becomes only
$A_{nbL}=1.4$. Regions near the centre of the galaxy are obscured by patchy
extinction clouds with $A_V$ as high as $\sim$40. In order to more fully
study these regions, multiband K and L band imaging will be important, to
disentangle the extinction patterns from the distribution patterns; even J and
H observations can suffer too much extinction in some of these regions to allow
reliable source detection. As we show, a nearly complete match can be made
between L objects and K objects down to limiting magnitudes of about 11,
suggesting that both bands penetrate sufficiently
to see as many sources as there are to be
seen.

Moreover we see an asymmetry consistent with bar models. Although the statistical weight
of our results is poor (we achieved less than half the survey observations in nbL that we
were hoping for) we show that this method - of surveying in K and L bands - does allow the
central regions of the Galaxy to be reached reliably in the plane.

To date, derivations of bar structure have depended on observations well out of the
plane due to the overwhelming problem of patchy extinction. We show that with suitably
penetrating wavelengths (K and L), direct evidence in the Galactic plane of dereddened
number count asymmetries can be found. Direct determination of the spatial
structure of the inner Galaxy is feasible with this technique.

\section{Acknowledgements}

The authors would like to thank the DENIS team for providing K band images in
the regions of interest. In particular, MU would like to thank Jean Borsenberger
and Francine Tanguy for their help in retrieving images from the DENIS database,
and his hosts Guy Simon at the Observatoire de Paris, and Alain Omont at the
IaP. MU acknowledges the financial support of the Particle Physics and Astronomy
Research Council.

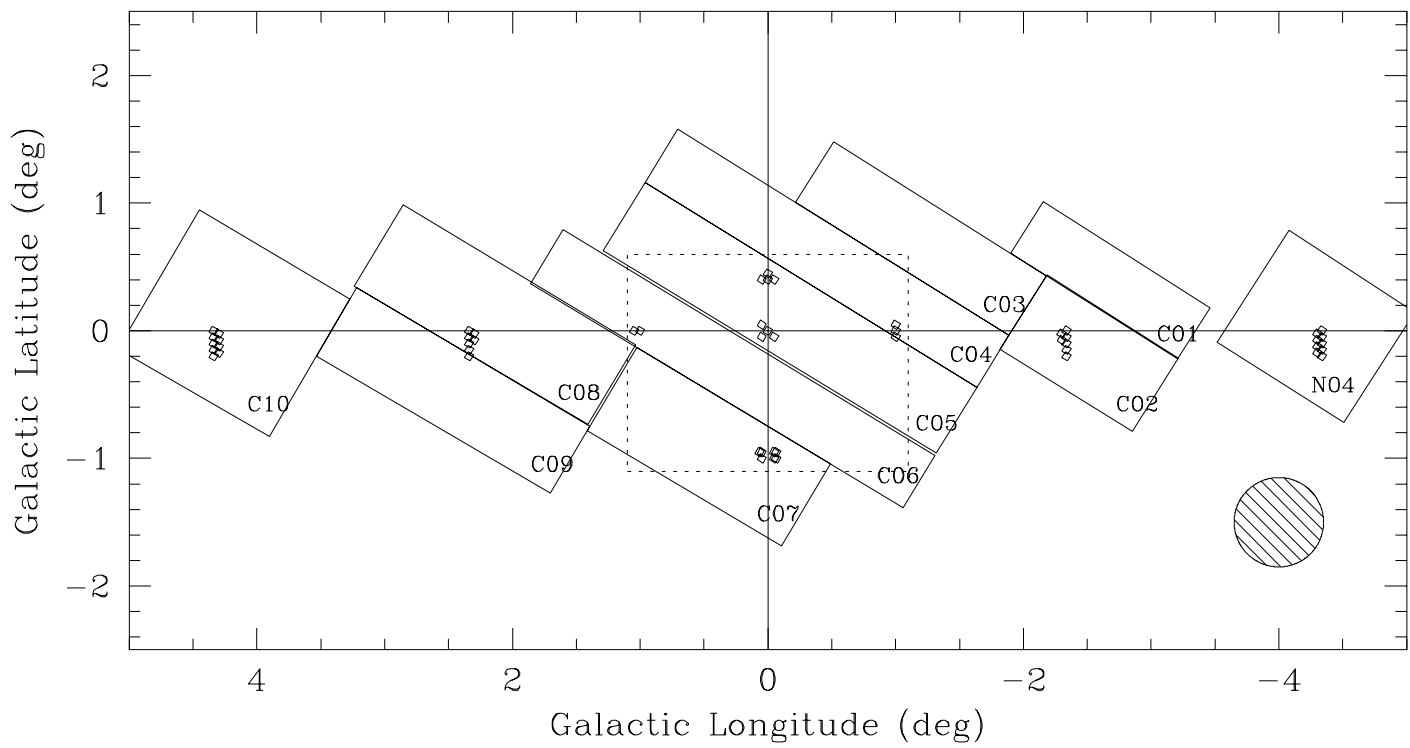
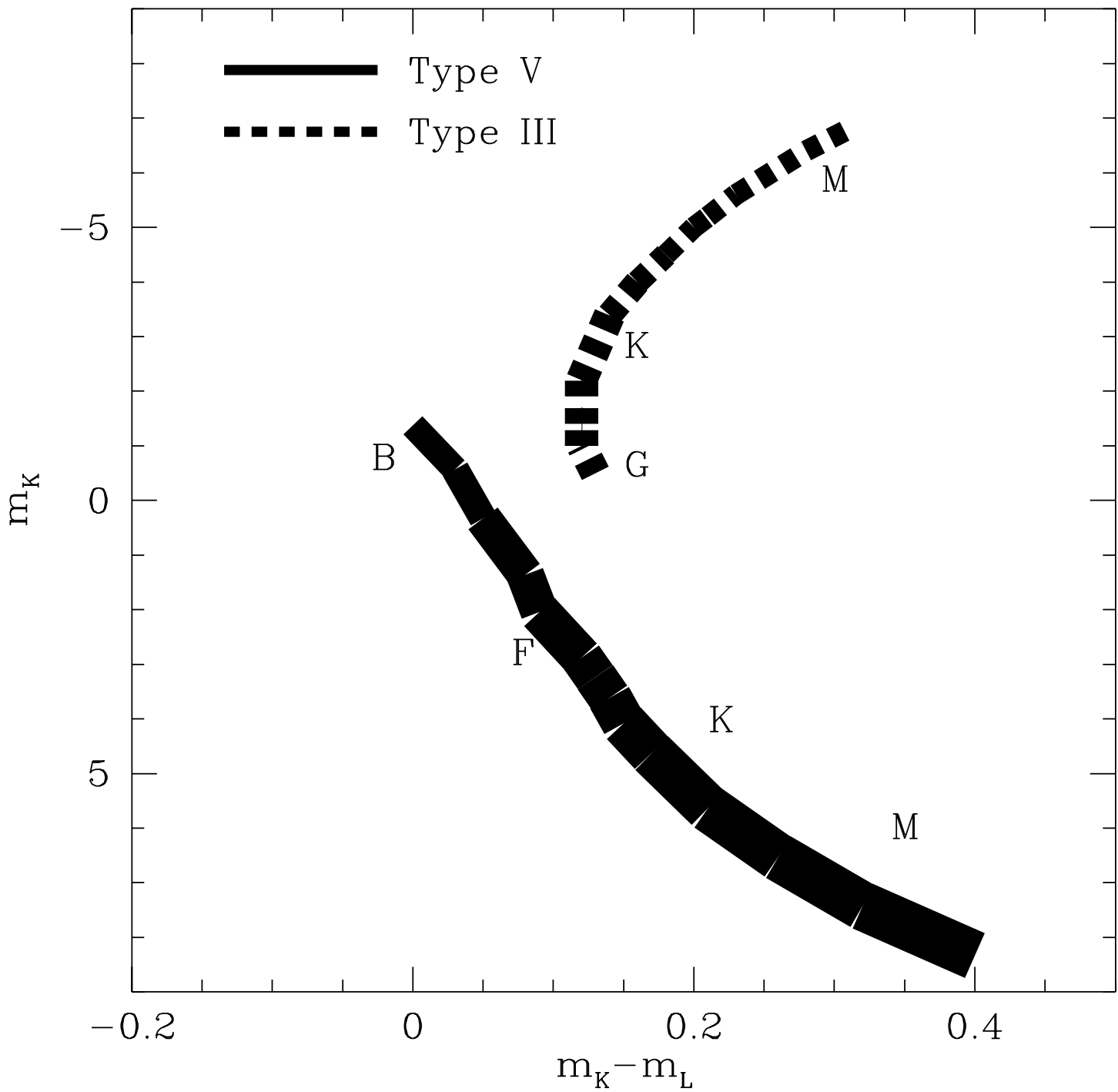
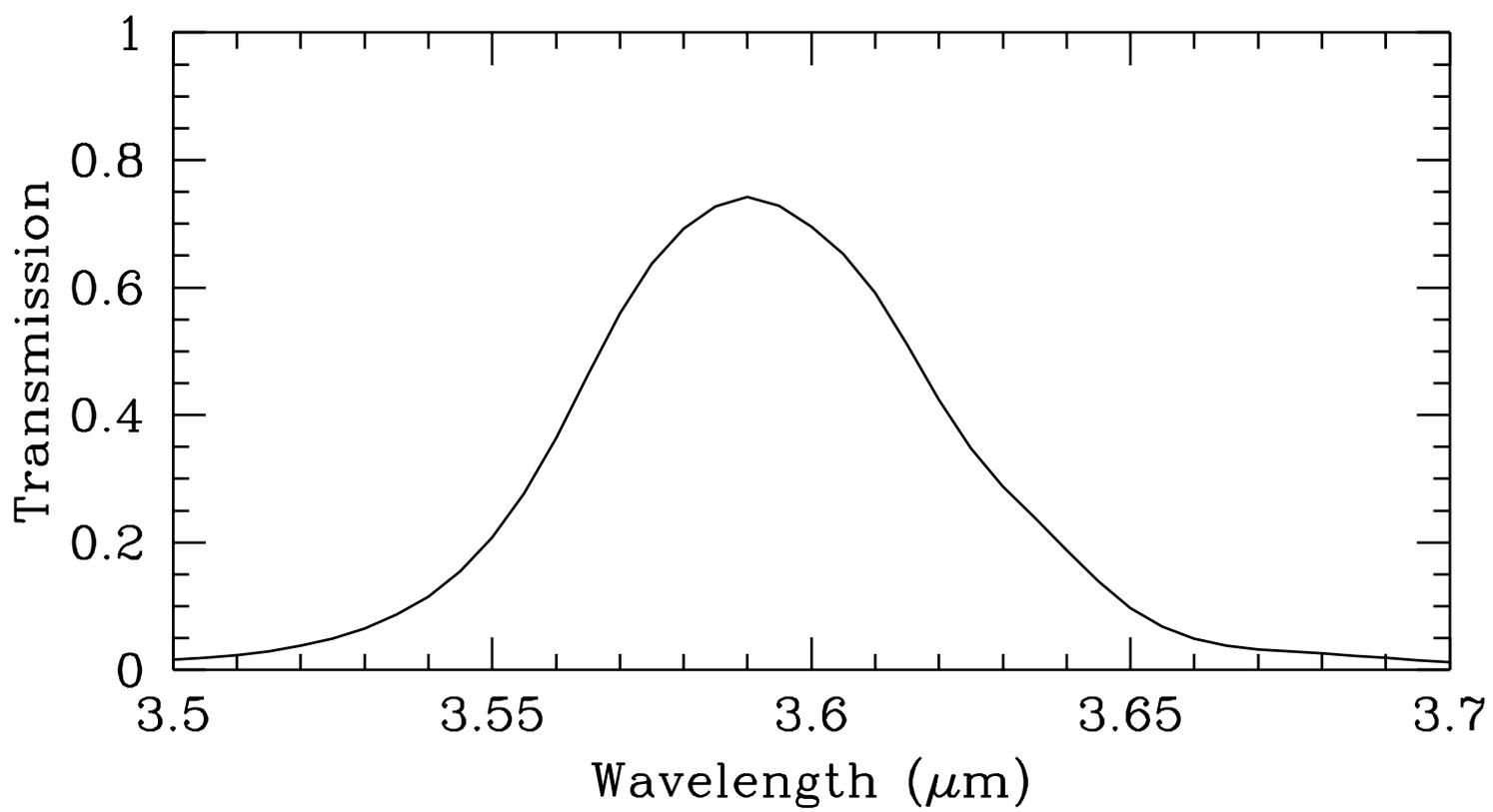
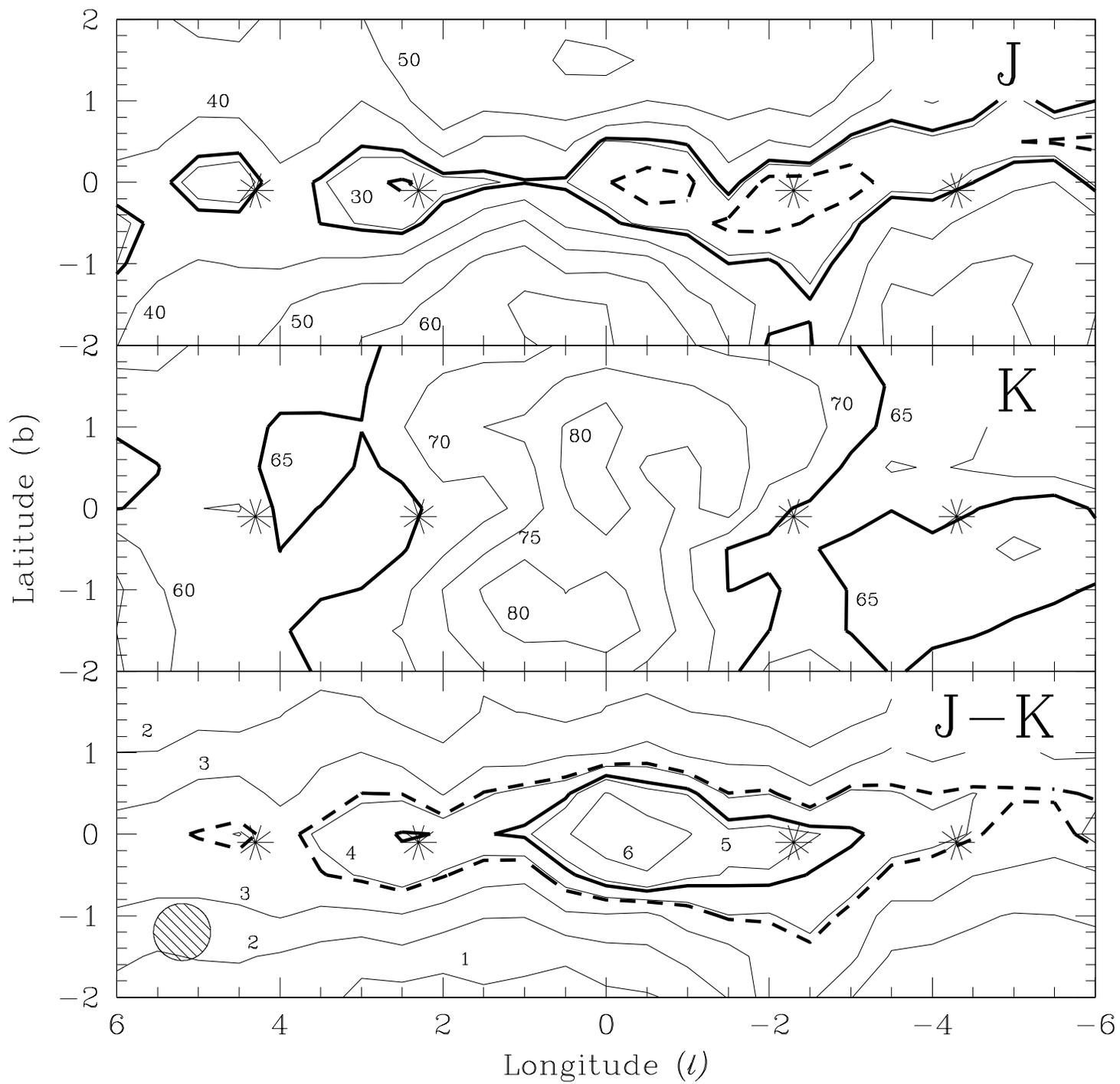
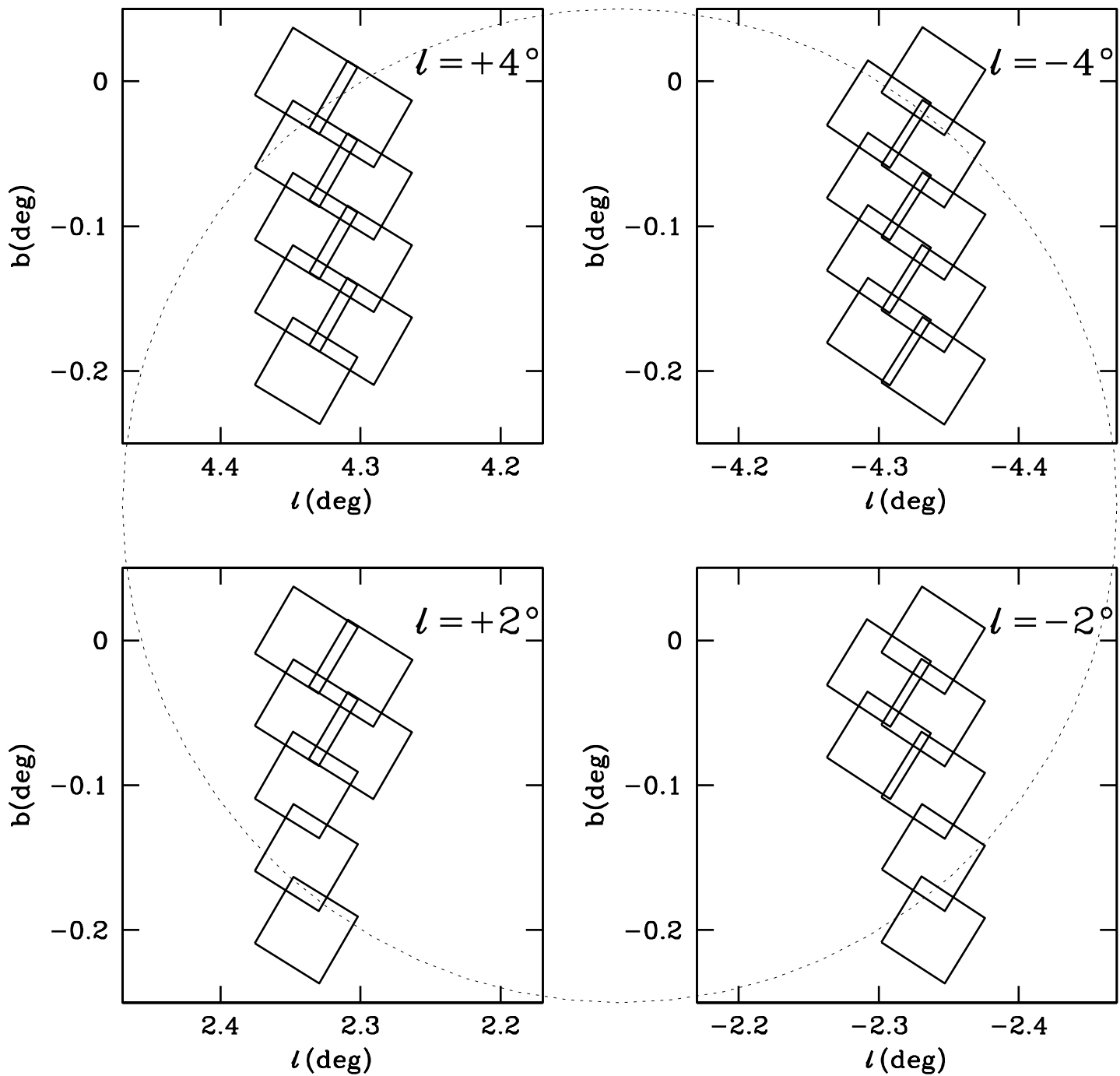
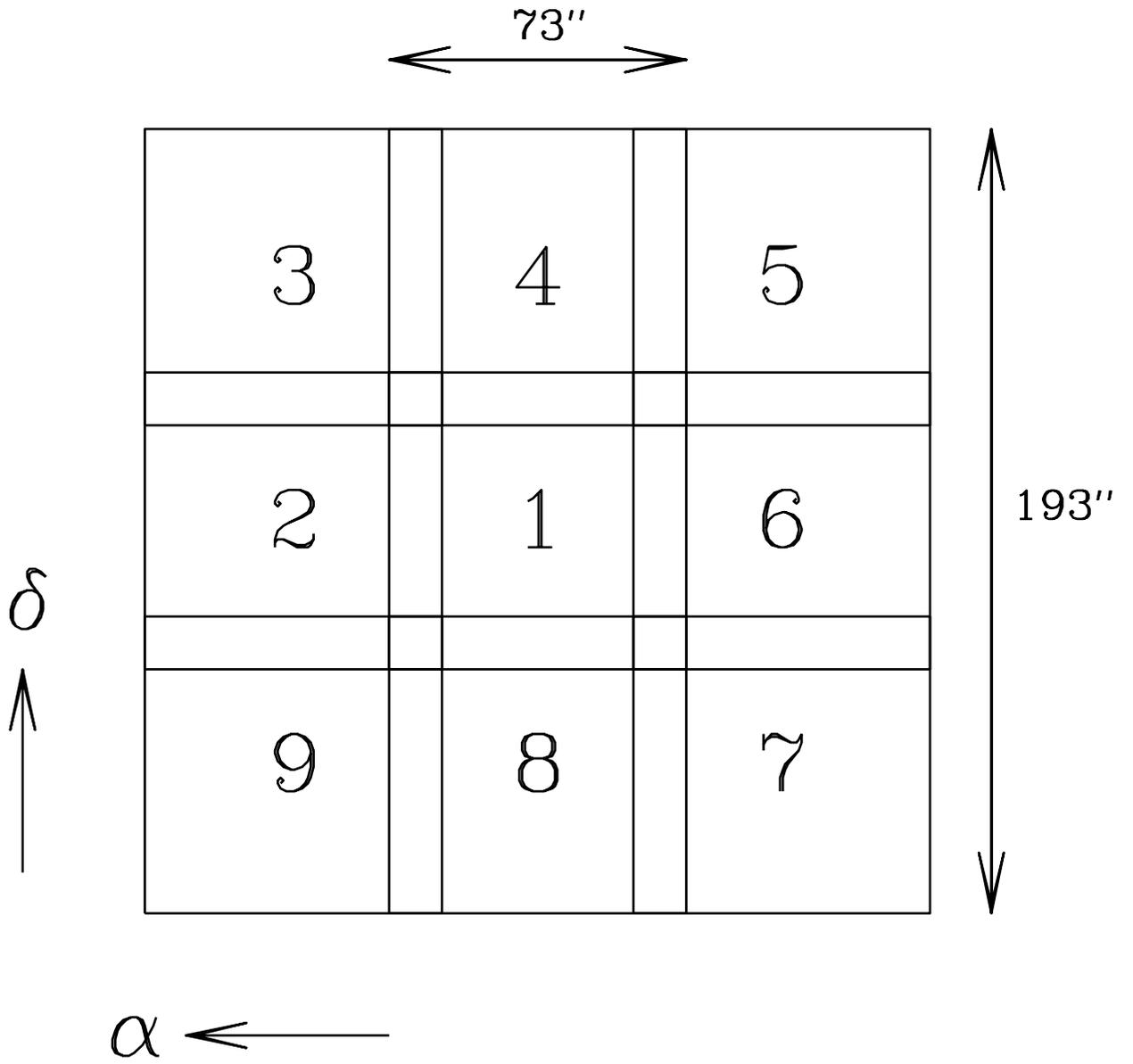
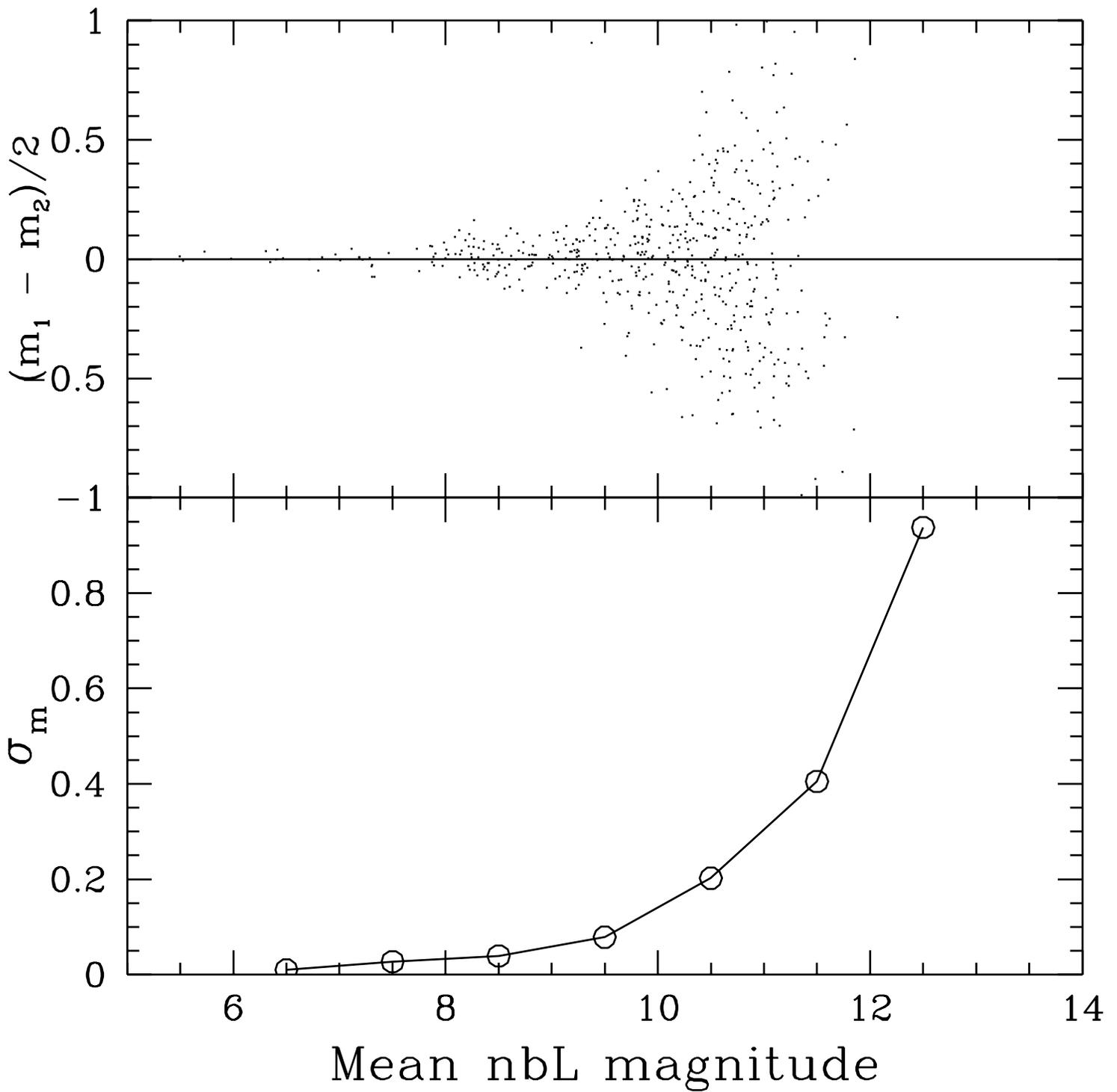
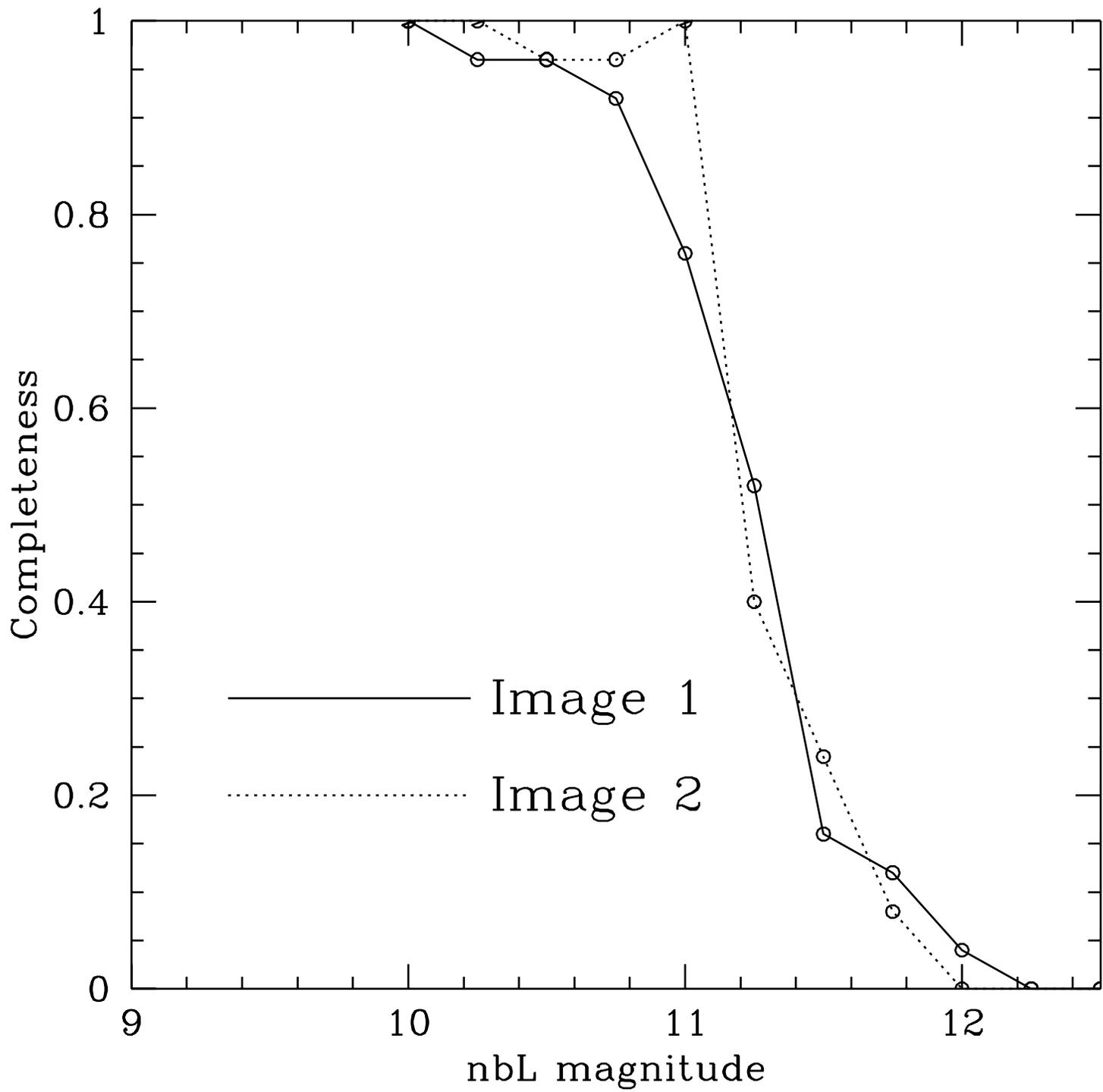
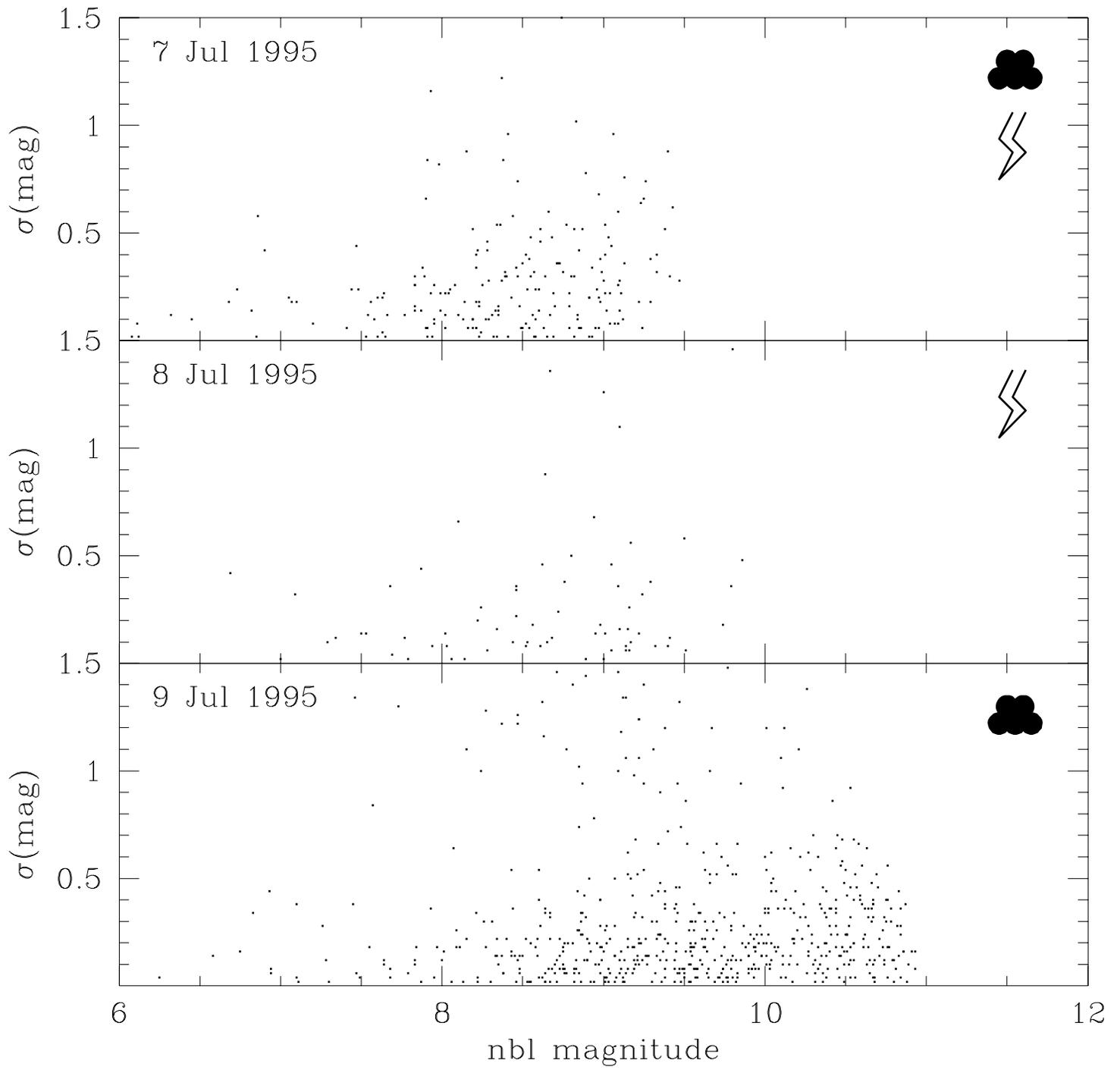
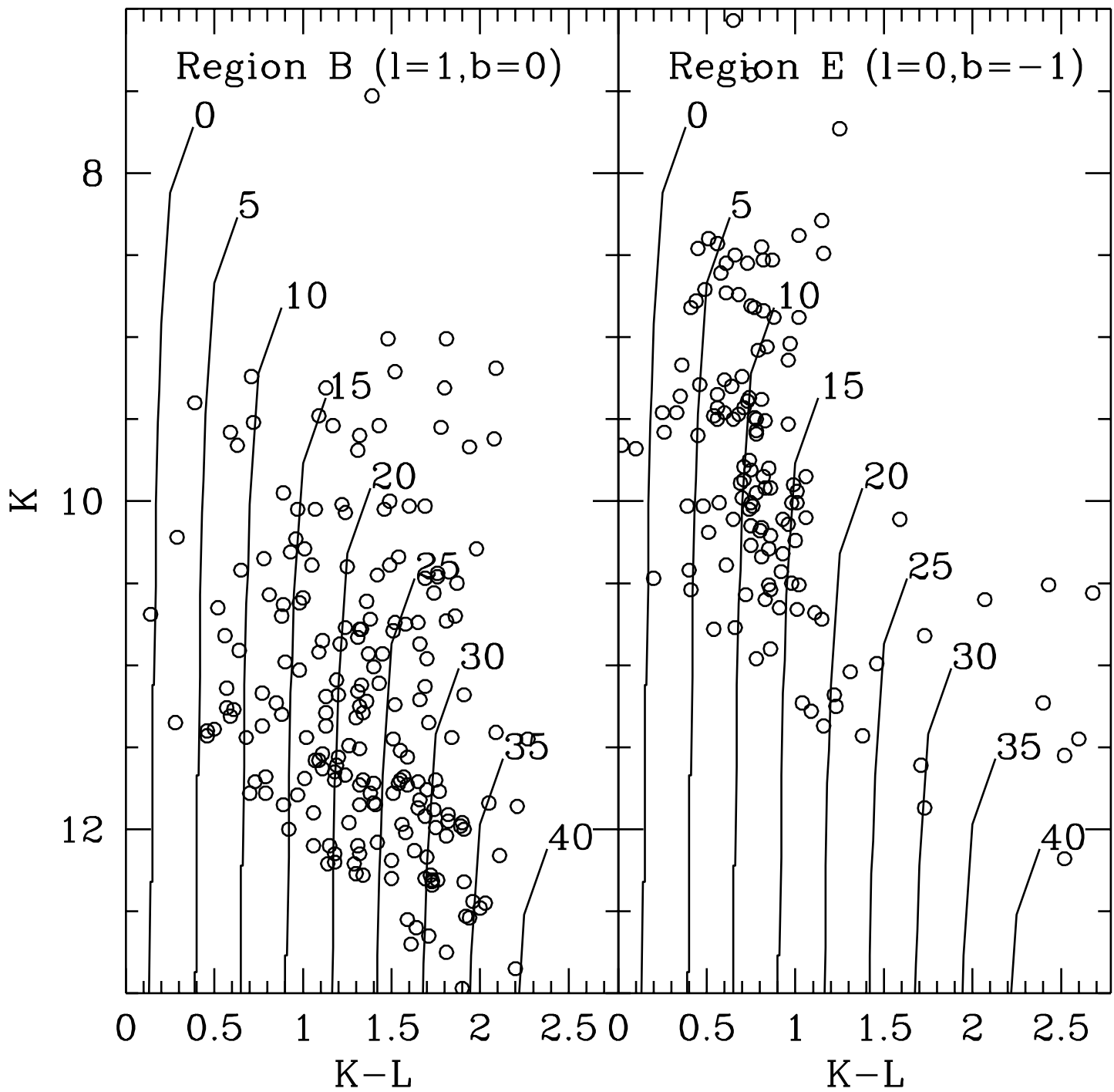
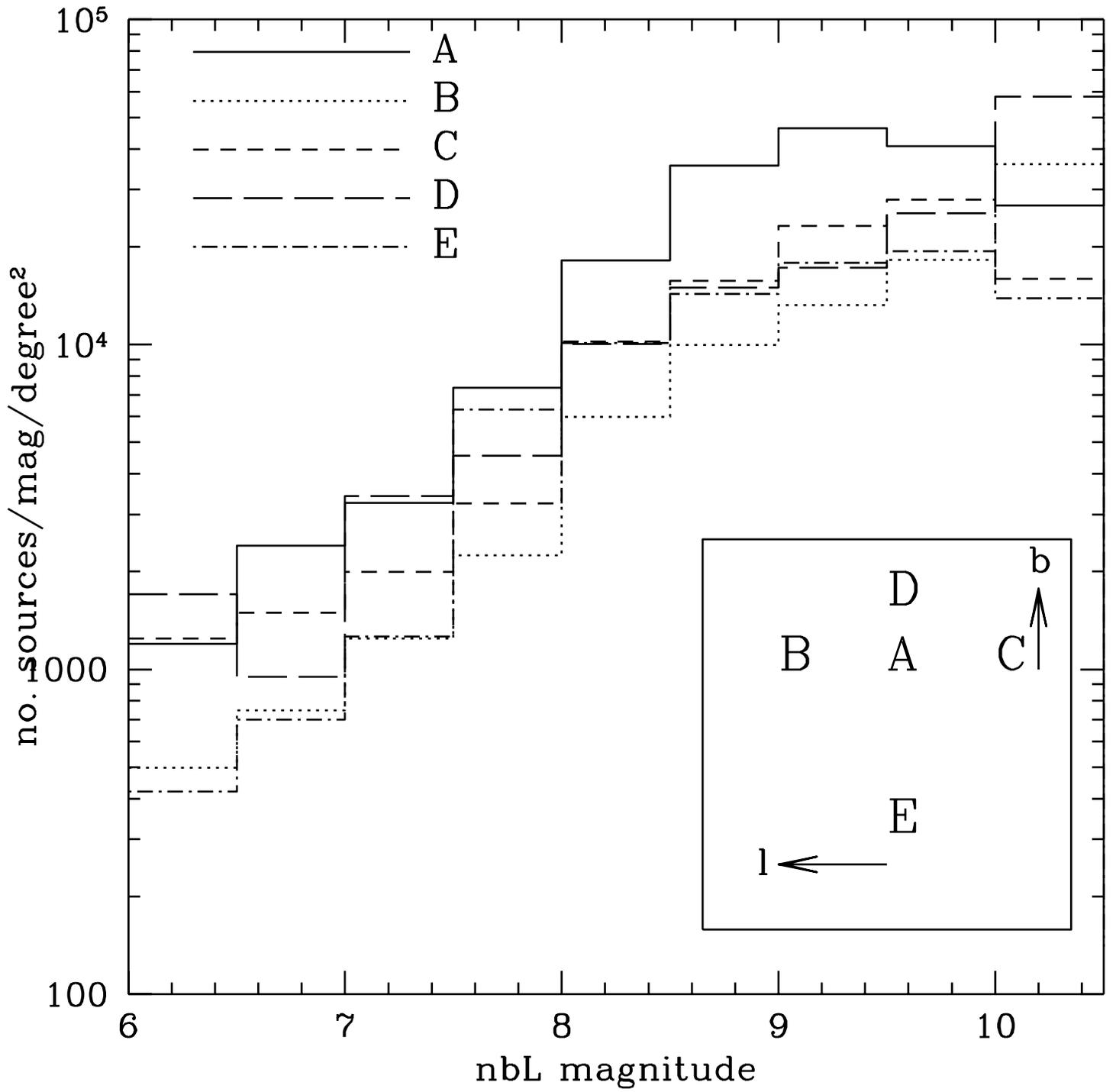
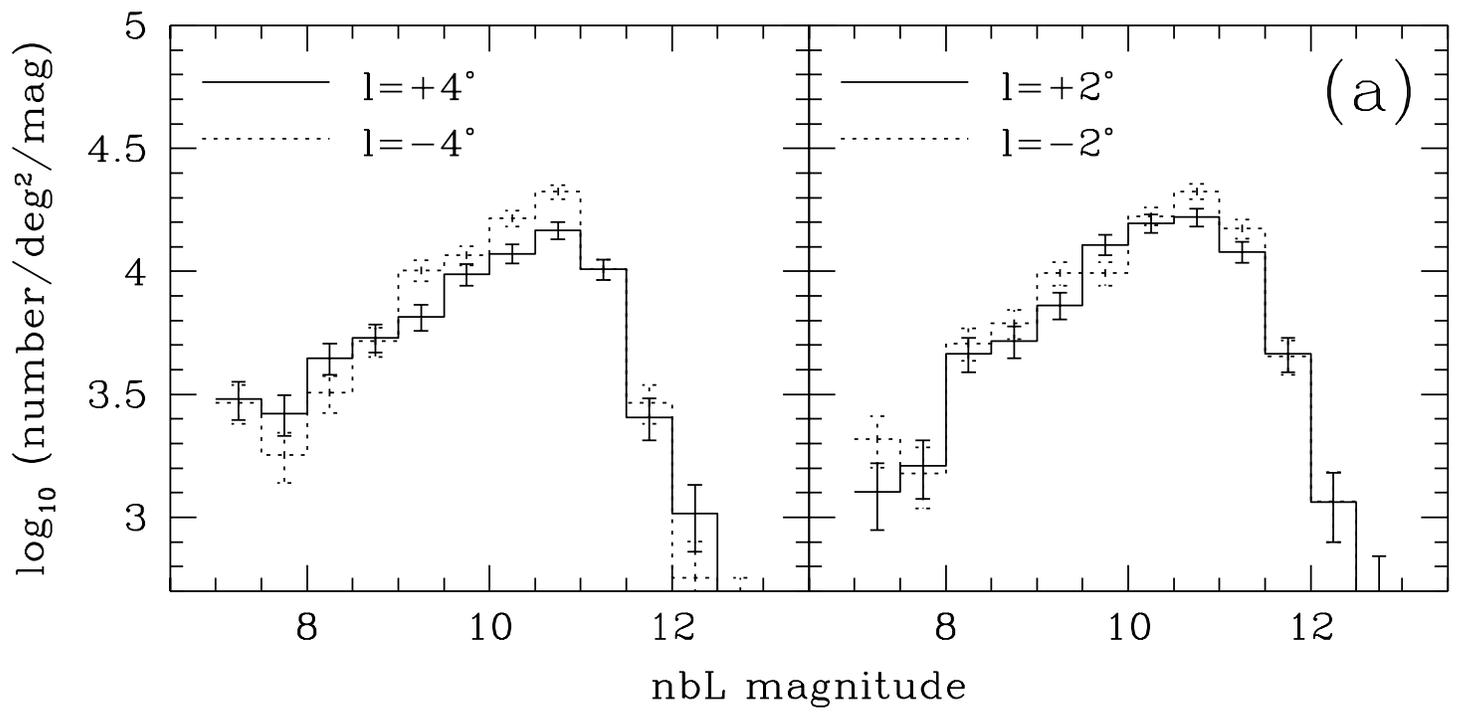
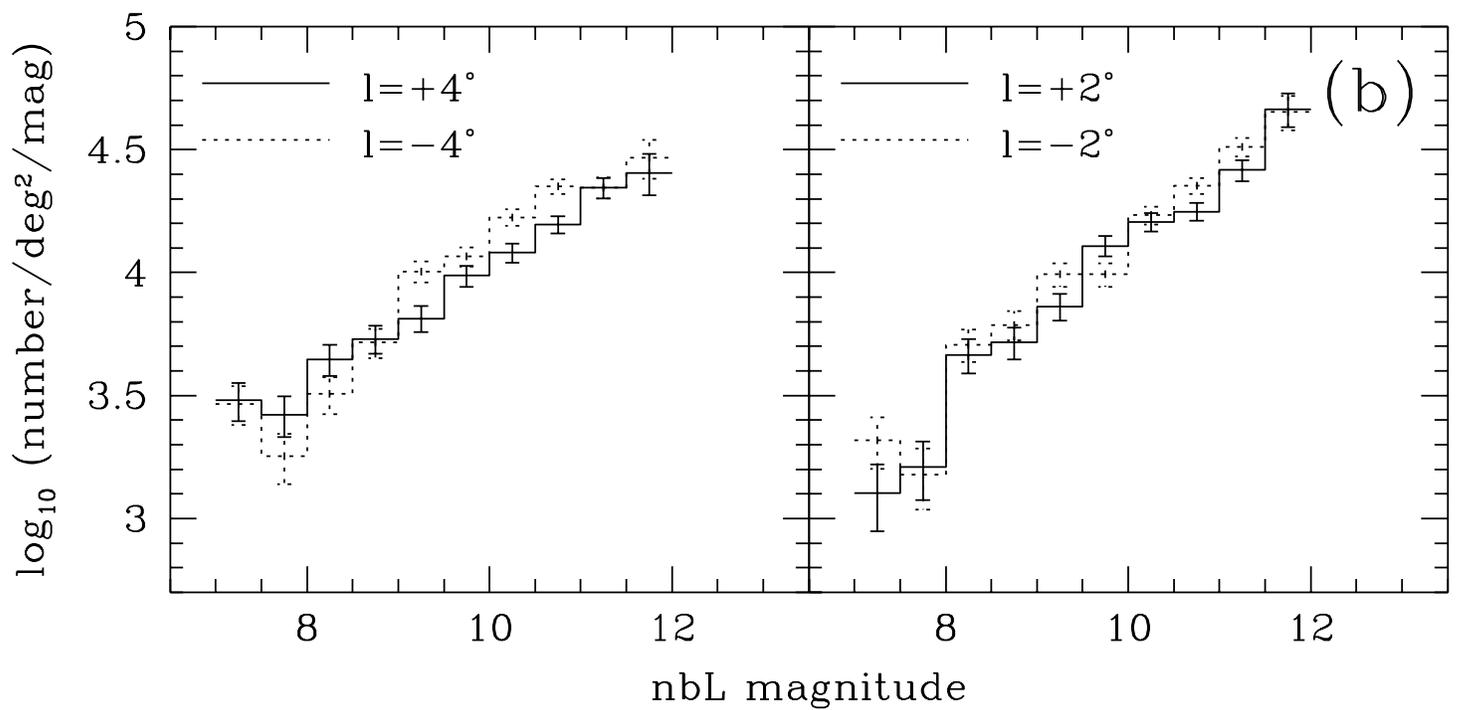
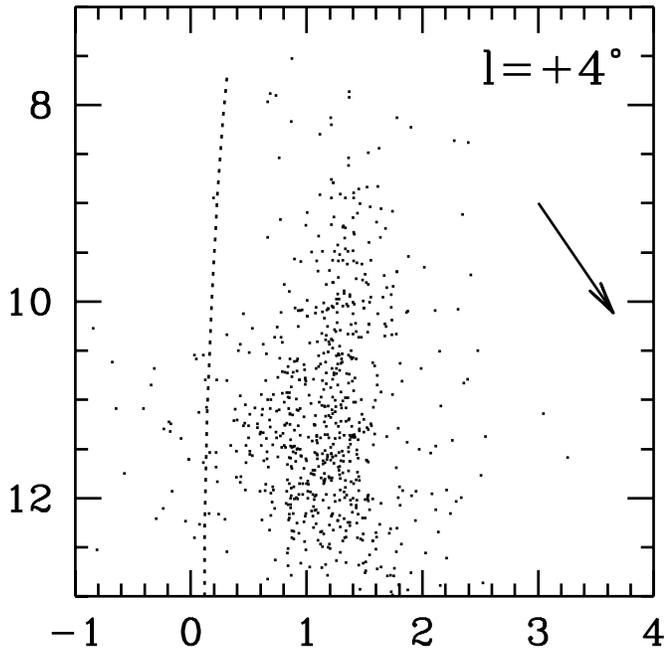
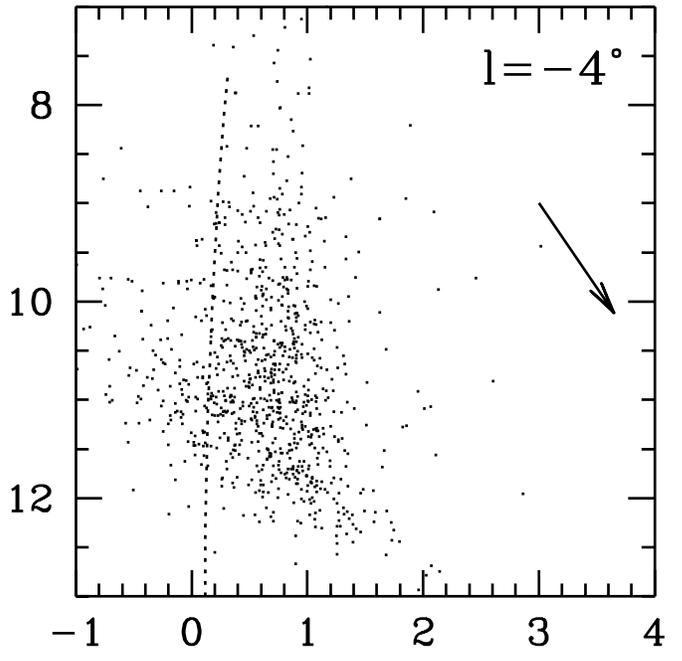
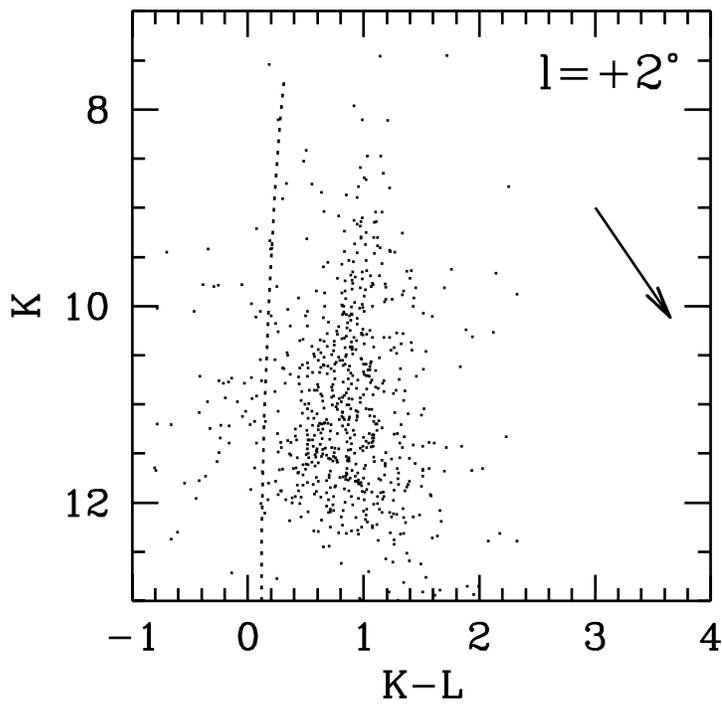
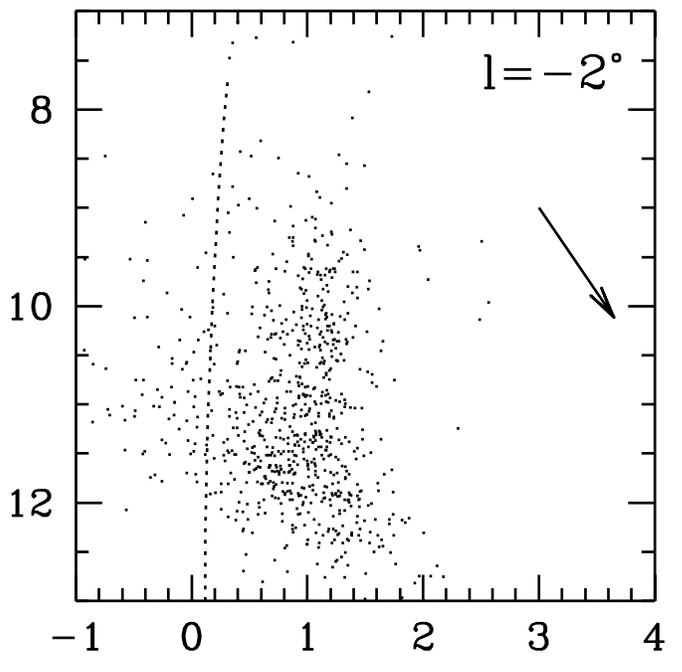
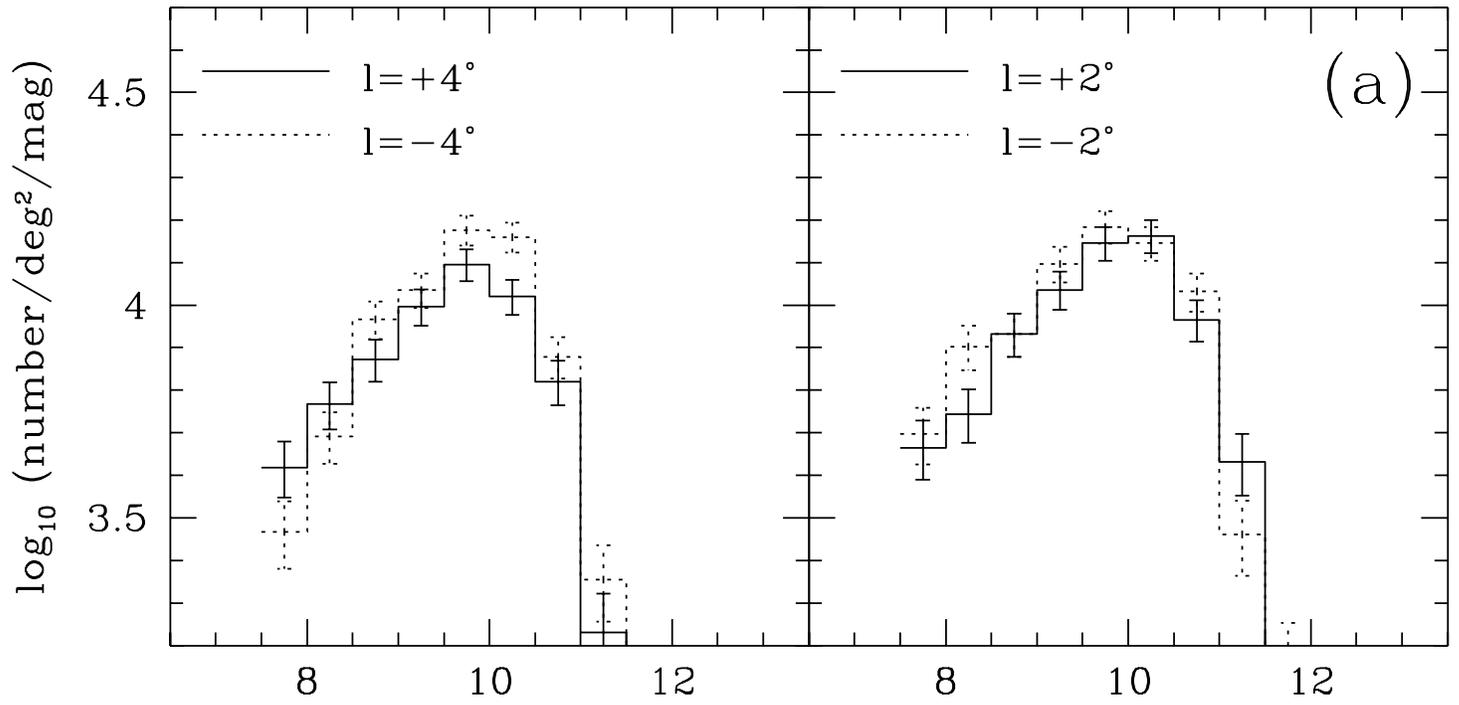
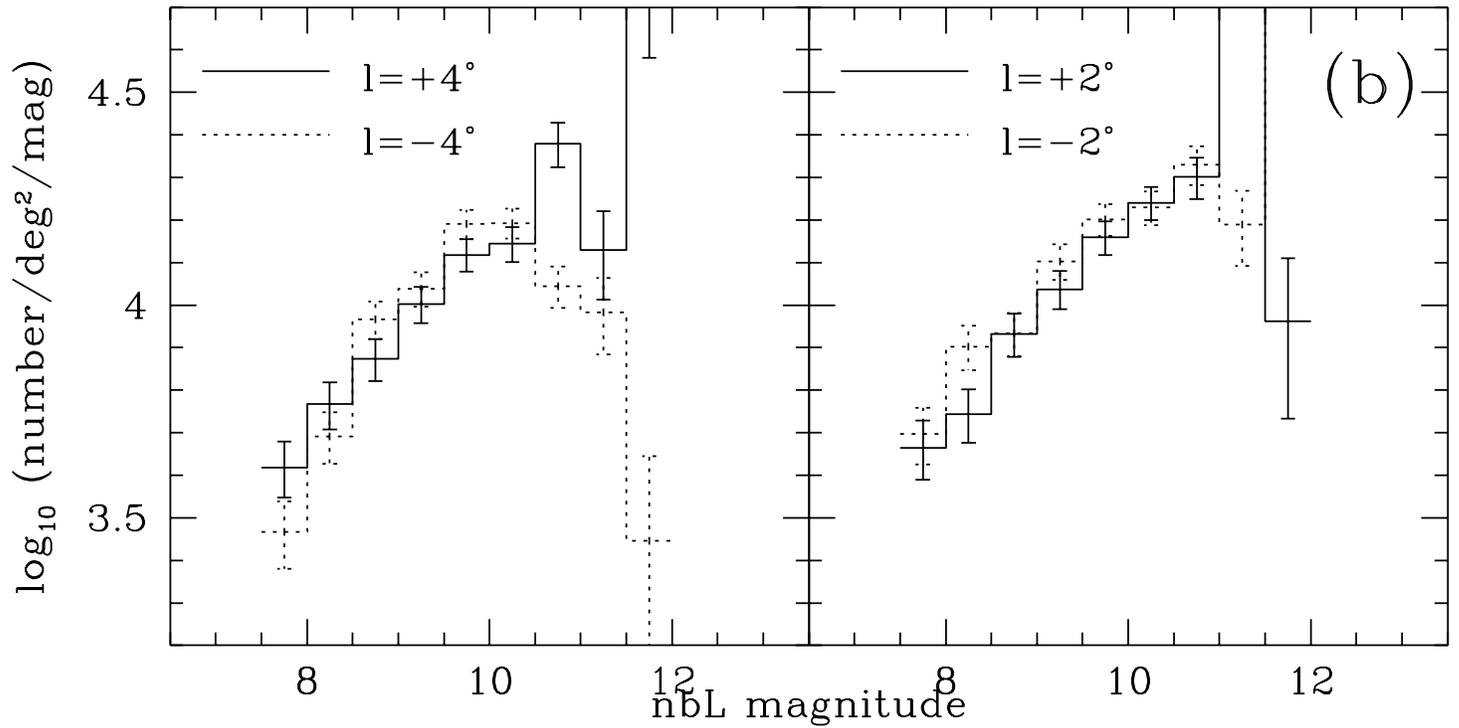
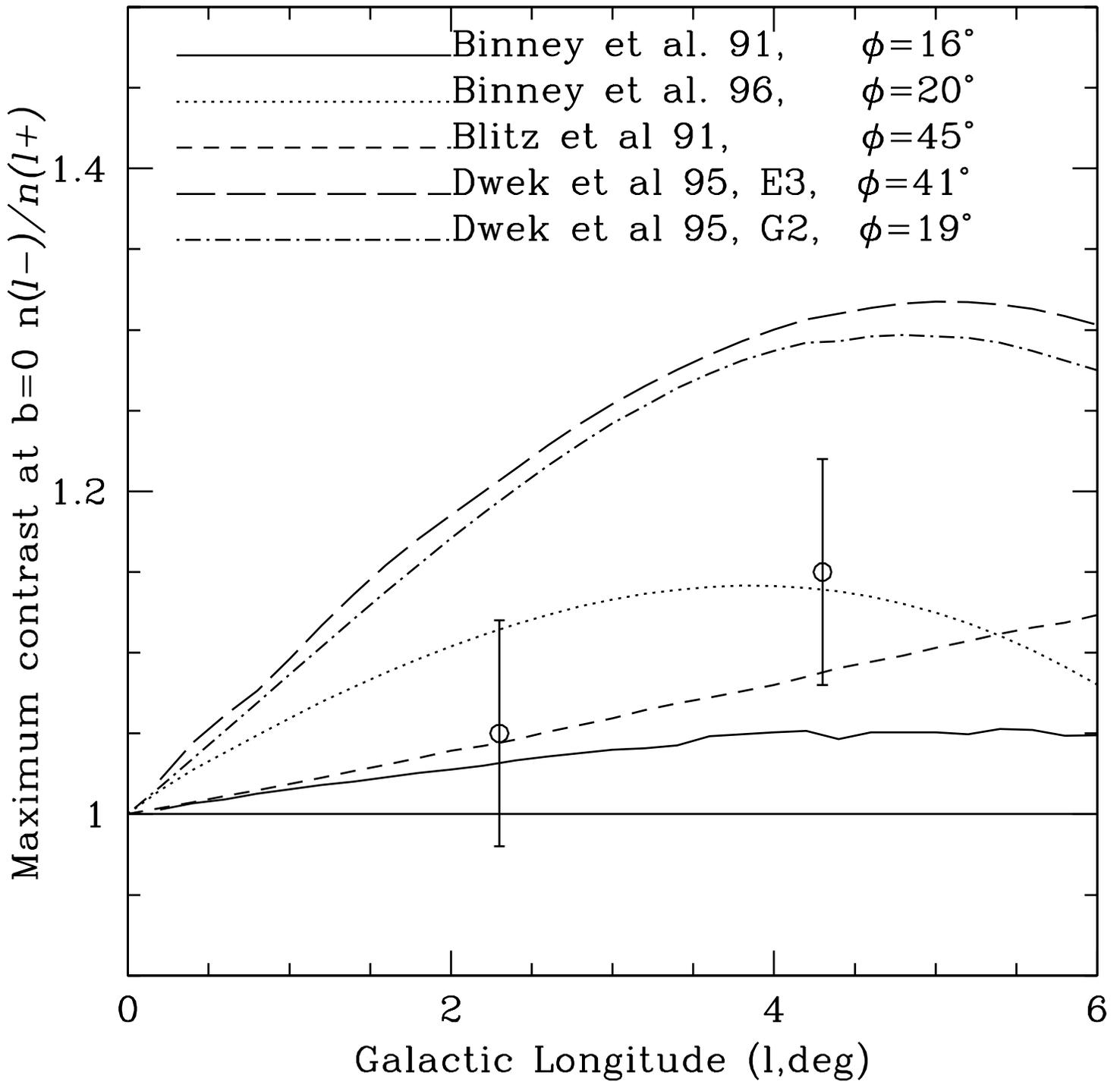
\end{document}